\documentclass{sigcomm-alternate}

\usepackage{listings}
\usepackage{ifpdf}
\ifpdf
\else
\fi

\usepackage{tabularx}
\usepackage{graphicx}
\usepackage{subfigure}
\usepackage[ruled,vlined]{algorithm2e}
\usepackage{enumitem}
\usepackage{float}
\usepackage{breakurl}
\usepackage[hyphens]{url}
\setlist{nolistsep,leftmargin=*}
\usepackage{balance}

\usepackage{breakurl}
\usepackage{color}
\usepackage{comment}
\usepackage{multirow}
\usepackage{url}
\usepackage[normalem]{ulem}

\usepackage{amssymb,amsmath,amsthm}

\numberwithin{subcase}{case}

\graphicspath{{figure/}}

\usepackage{leading}
\leading{12.08pt}

\newcommand{\matchaction}{match\raise.30ex\hbox{\smaller{+}}{\allowbreak}action\xspace}
\newcommand{\MatchAction}{Match\raise.30ex\hbox{\smaller{+}}{\allowbreak}Action\xspace}
\newcommand{\Matchaction}{Match\raise.30ex\hbox{\smaller{+}}{\allowbreak}action\xspace}

\begin{document}
\date{}

\title{P4: Programming Protocol-Independent \\Packet Processors}
%\numberofauthors{1}
\author{
\alignauthor
    Pat~Bosshart\textsuperscript{\dag},
    Dan~Daly\textsuperscript{*},
    Glen~Gibb\textsuperscript{\dag},
    Martin~Izzard\textsuperscript{\dag},
    Nick~McKeown\textsuperscript{\ddag},
    Jennifer~Rexford\textsuperscript{**},
    Cole~Schlesinger\textsuperscript{**},
    Dan~Talayco\textsuperscript{\dag},
    Amin~Vahdat\textsuperscript{\P},
    George~Varghese\textsuperscript{\S},
    David~Walker\textsuperscript{**}\\
    \affaddr{
      \textsuperscript{\dag}Barefoot Networks\quad
      \textsuperscript{*}Intel\quad
      \textsuperscript{\ddag}Stanford University\quad
      \textsuperscript{**}Princeton University\quad
      \textsuperscript{\P}Google\quad
      \textsuperscript{\S}Microsoft Research
    }\\
%    \email{
%      \{bosshart, martin, dan\}@barefootnetworks.com\;
%     dan.daly@intel.com\;
%      nickm@stanford.edu\;
%      \{jrex, dpw\}@cs.princeton.edu
%      vahdat@google.com\;
%      varghese@microsoft.com\;
%    }
}

\maketitle

\begin{abstract}
P4 is a high-level language for programming protocol-inde\-pendent 
packet processors.  P4 works in conjunction with SDN control protocols
like OpenFlow. In its current form, OpenFlow explicitly specifies
protocol headers on which it operates.  This set has grown from 12 to 41
fields in a few years, increasing the complexity of the specification while
still not providing the flexibility to add new headers.
In this paper we propose P4 as a strawman proposal for how OpenFlow
should evolve in the future.  We have three goals: (1)
Reconfigurability in the field: Programmers should be able to change
the way switches process packets once they are deployed.
(2) Protocol independence: Switches should not be tied to any
specific network protocols. (3) Target independence: Programmers
should be able to describe packet-processing functionality independently
of the specifics of the underlying hardware.
As an example, we describe how to use P4 to configure a
switch to add a new hierarchical label.
\end{abstract}

\balance

\section{Introduction}
\label{intro}
%%%
%%% SDN is awesome but the number of headers is proliferating
%%%
Software-Defined Networking (SDN) gives operators programmatic control over their networks. In SDN, the control plane is physically separate from the forwarding plane, and one control plane controls multiple forwarding devices. While forwarding devices could be programmed in many ways, having a common, open, vendor-agnostic interface (like OpenFlow) enables a control plane to control forwarding devices from different hardware and software vendors.

\restylefloat{table}
\begin{table}[H]
\begin{center}
\small
\begin{tabular}{|c|l|l|} \hline
\multicolumn{1}{|c}{\bf Version} &
  \multicolumn{1}{|c}{\bf Date} &
  \multicolumn{1}{|c|}{\bf Header Fields} \\\hline
OF 1.0 & Dec 2009 & 12 fields (Ethernet, TCP/IPv4) \\
OF 1.1 & Feb 2011 & 15 fields (MPLS, inter-table metadata) \\
OF 1.2 & Dec 2011 & 36 fields (ARP, ICMP, IPv6, etc.) \\
OF 1.3 & Jun 2012 & 40 fields \\
OF 1.4 & Oct 2013 & 41 fields \\ \hline
\end{tabular}
\end{center}
\caption{Fields recognized by the OpenFlow standard}
\label{tab:openflow}
\end{table}

The OpenFlow interface started simple, with the abstraction of a single table of rules that could match packets on a dozen header fields (e.g., MAC addresses, IP addresses, protocol, TCP/UDP port numbers, etc.).  Over the past five years, the specification has grown increasingly more complicated (see Table~\ref{tab:openflow}), with many more header fields and multiple stages of rule tables, to allow switches to expose more of their capabilities to the controller.

%%%
%%%  Stopping the chaos
%%%
The proliferation of new header fields shows no signs of stopping.  For
example, data-center network operators increasingly want to apply new forms of packet encapsulation (e.g., NVGRE, VXLAN, and STT), for which they resort to deploying software switches that are easier to extend with new functionality.  Rather than repeatedly extending the OpenFlow specification, we argue that future switches should support flexible mechanisms for parsing packets and
matching header fields, allowing controller applications to leverage these capabilities through a common, open interface (i.e., a new ``OpenFlow 2.0'' API).  Such a general, extensible approach would be simpler, more elegant, and more future-proof than today's OpenFlow 1.x standard.

\begin{figure}[h]
\centering
\includegraphics[width=2.5in]{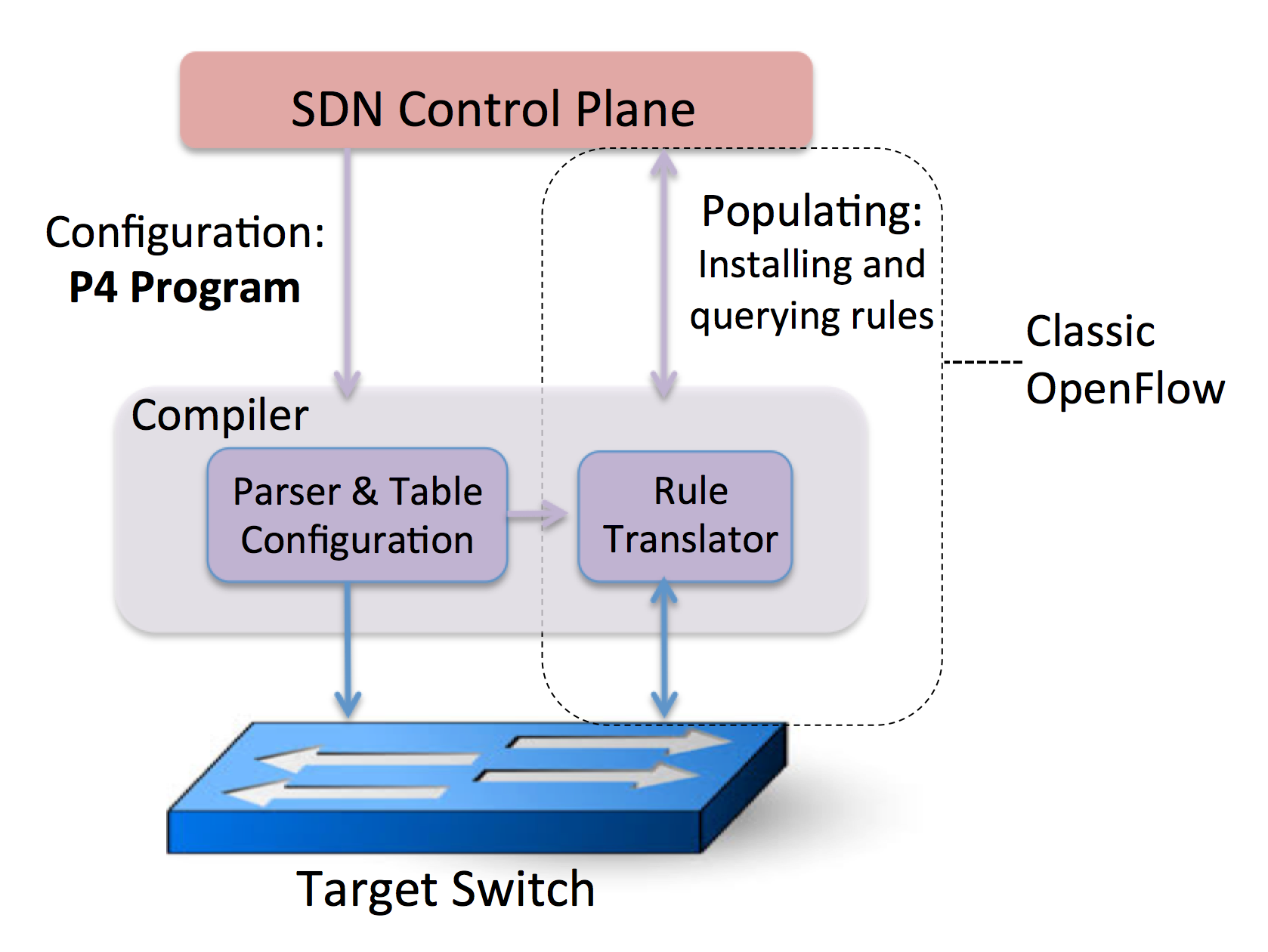}
\caption{P4 is a language to configure switches.}
\label{fig:approach}
\end{figure}

%%%
%%% Need a programming interface
%%%
Recent chip designs demonstrate that such flexibility can be achieved in custom ASICs at terabit speeds~\cite{kangaroo,forwarding-metamorphosis, intel_fm6000}.
Programming this new generation of switch chips is far from easy.  Each chip
has its own low-level interface, akin to microcode programming.  In this
paper, we sketch the design of a higher-level language for \uline{P}rogramming
\uline{P}rotocol-independent \uline{P}acket \uline{P}rocessors (P4).  Figure~\ref{fig:approach} shows the relationship between P4---used to configure a switch, telling it how packets are to be processed---and existing APIs (such as OpenFlow) that are designed to populate the forwarding tables in fixed function switches.
P4 raises the level of abstraction for programming the network, and can serve as a general interface between the controller and the switches.  That is, we believe that future generations of OpenFlow should allow the controller to \emph{tell the switch how to operate}, rather than be constrained by a fixed switch design. The key challenge is to find a ``sweet spot'' that balances the need for expressiveness with the ease of implementation across a wide range of hardware and software switches.  In designing P4, we have three main goals:

\begin{itemize}
\item{\textbf{Reconfigurability.}} The controller should be able to redefine the packet parsing and processing
in the field.
\item{\textbf{Protocol independence.}}  The switch should not be tied to specific packet formats. Instead, the controller should be able to specify (i) a packet parser for extracting header fields with particular names and types and (ii) a collection of typed \matchaction tables that process these headers.
\item{\textbf{Target independence.}} Just as a C programmer does not need to know the specifics of the underlying CPU, the controller programmer should not need to know the details of the underlying switch.  Instead, a compiler should take the switch's capabilities into account when turning a target-independent description (written in P4) into a target-{\em de}pendent program (used to configure the switch).
\end{itemize}

%%%
%%% Roadmap
%%%
\noindent
The outline of the paper is as follows. We begin by introducing an abstract
switch forwarding model. Next, we explain the need for a new language to
describe protocol-independent packet processing. We then present a simple
motivating example where a network operator wants to support a new
packet-header field and process packets in multiple stages.  We use this to
explore how the P4 program specifies headers, the packet parser, the multiple
\matchaction tables, and the control flow through these tables.  Finally, we
discuss how a compiler can map P4 programs to target switches.

\textbf{Related work.}  
%Several papers laid the groundwork for our proposal. Most notably 
In 2011, Yadav et al.~\cite{yadav} proposed an abstract forwarding model for OpenFlow, but with less emphasis on a compiler. Kangaroo~\cite{kangaroo} introduced the notion of programmable parsing. Recently, Song~\cite{POF} proposed protocol-oblivious forwarding which shares our goal of protocol independence, but is targeted more towards network processors. The ONF introduced table typing patterns to express the matching capabilities of switches~\cite{FAWG-doc}.  Recent work on NOSIX~\cite{nosix} shares our goal of flexible specification of \matchaction tables, but does not consider protocol-independence or propose a language for specifying the parser, tables, and control flow.  Other recent work proposes a programmatic interface to the data plane for monitoring, congestion control, and queue management~\cite{tpp,silver}.  The Click modular router~\cite{click} supports flexible packet processing in software, but does not map programs to a variety of target hardware switches.

\section{Abstract Forwarding Model}
\label{model}

\begin{figure}[h]
\centering
\includegraphics[width=\columnwidth]{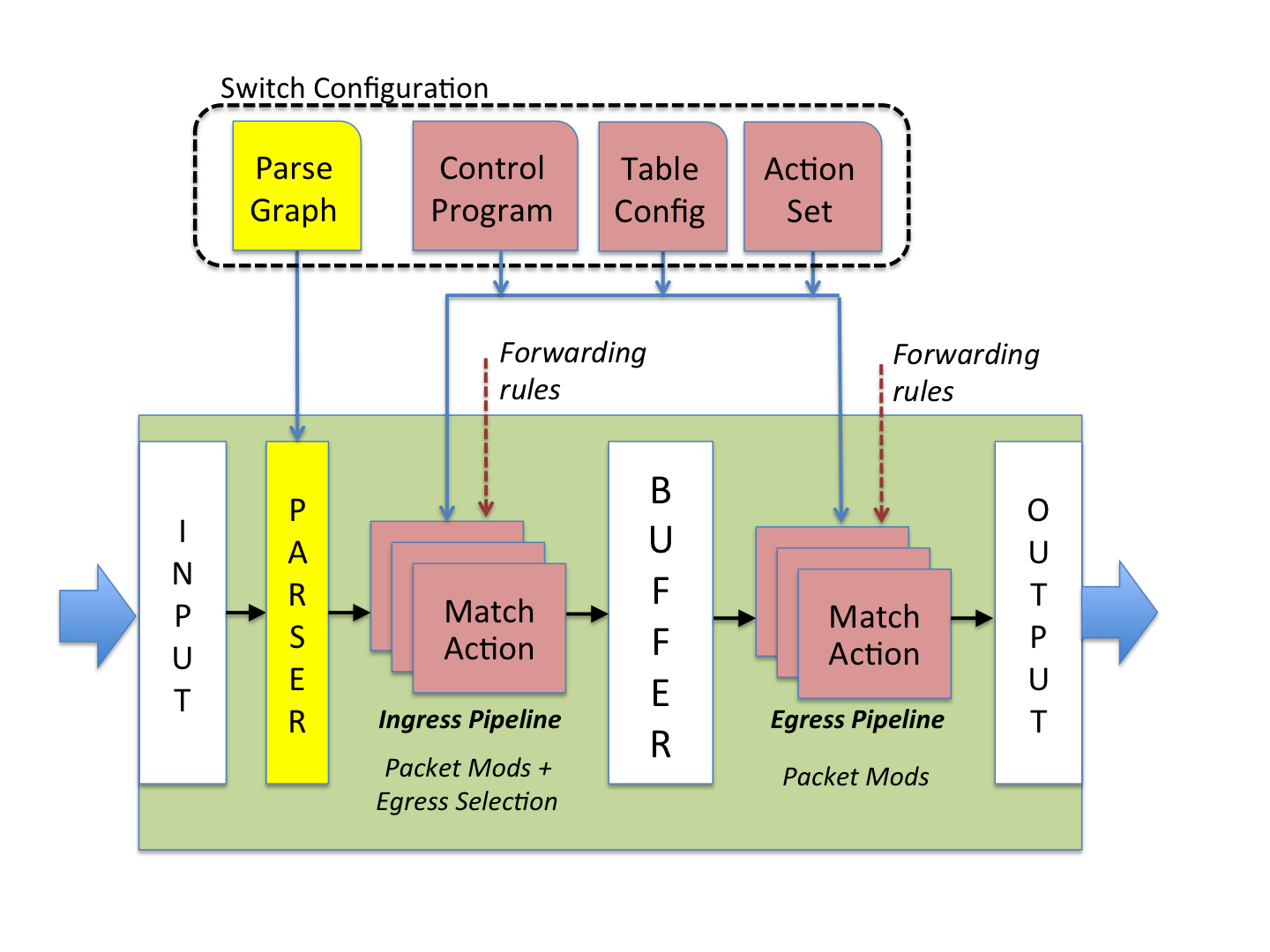}
\caption{The abstract forwarding model.}
\label{fig:model}
\end{figure}

In our abstract model (Fig.~\ref{fig:model}), switches forward packets via a programmable parser followed by multiple stages of \matchaction, arranged in series, parallel, or a combination of both. Derived from OpenFlow, our model makes three generalizations. First, OpenFlow assumes a fixed parser, whereas our model supports a programmable parser to allow new headers to be defined. Second, OpenFlow assumes the \matchaction stages are in series, whereas in our model they can be in parallel or in series. Third, our model assumes that actions are composed from protocol-independent primitives supported by the switch.

Our abstract model generalizes how packets are processed in different forwarding devices (e.g., Ethernet switches, routers, load-balancers)\ and by different technologies (e.g., fixed-function switch ASICs, NPUs, reconfigurable switches, software switches, FPGAs). This allows us to devise a common language (P4) to represent how packets are processed in terms of our common abstract model. Hence, programmers can create target-independent programs that a compiler can map to a variety of different forwarding devices, ranging from relatively slow software switches to the fastest ASIC-based switches.

The forwarding model is controlled by two types of operations:
Configure and Populate.  {\em Configure} operations program the
parser, set the order of \matchaction stages, and specify the
header fields processed by each stage. Configuration determines which
protocols are supported and how the switch may process packets.
{\em  Populate} operations add (and remove) entries to the
\matchaction tables that were specified during configuration.
Population determines the policy applied to packets at any given
time.

For the purposes of this paper, we assume that configuration and
population are two distinct phases. In particular, the switch need not
process packets during configuration.  However, we expect
implementations will allow packet processing during partial or
full reconfiguration enabling upgrades with no downtime. Our model
deliberately allows for, and encourages, reconfiguration that does not
interrupt forwarding.

Clearly, the configuration phase has little meaning in fixed-function ASIC switches; for this type of switch, the compiler's job is to simply check if the chip can support the P4 program. Instead, our goal is to capture the general trend towards fast reconfigurable packet-processing pipelines, as described in~\cite{forwarding-metamorphosis, intel_fm6000}.

Arriving packets are first handled by the parser. The packet body is assumed to be
buffered separately, and unavailable for matching. The parser recognizes and
extracts fields from the header, and thus defines the protocols
supported by the switch.  The model makes
no assumptions about the meaning of protocol headers, only that the parsed representation defines a collection of
fields on which matching and actions operate.

The extracted header fields are then passed to the \matchaction tables.  The \matchaction tables are divided between ingress and
egress.  While both may modify the packet header,
ingress \matchaction determines the egress port(s) and
determines the queue into which the packet is placed. Based on ingress
processing, the packet may be forwarded, replicated (for multicast,
span, or to the control plane), dropped, or trigger flow
control. Egress \matchaction performs per-instance modifications to
the packet header -- e.g., for multicast copies.  Action tables (counters, policers, etc.) can be associated with a flow to track frame-to-frame state.

Packets can carry additional information between stages, called {\em metadata}, which is treated identically to packet header fields.  Some examples of metadata include the ingress port,
the transmit destination and queue, a timestamp that can be used for packet scheduling, and data passed from table-to-table that does not involve changing the parsed representation of
the packet such as a virtual network identifier.

Queueing disciplines are handled in the same way as the current OpenFlow: an action maps a packet to a queue, which is configured to receive a particular service discipline. The service discipline (e.g., minimum rate, DRR) is chosen as part of the switch configuration.

Although beyond the scope of this paper, action primitives can be added to allow the programmer to implement new or existing congestion control protocols. For example, the switch might be programmed to set the ECN bit based on novel conditions, or it might implement a  proprietary congestion control mechanism using \matchaction tables.

\section{A Programming Language}
\label{prog-lang}

We use the abstract forwarding model to define a language to express how a switch is to be configured and how packets are to be processed. This paper's main goal is to propose the P4 programming language.  However, we recognize that many languages are possible, and they will likely share the common characteristics we describe here. For example, the language needs a way to express how the parser is programmed so that the parser knows what packet formats to expect; hence a programmer needs a way to declare what header types are possible.  As an example, the programmer could specify the format of an IPv4 header and what headers may legally follow the IP header. This motivates defining parsing in P4 by declaring legal header types. Similarly, the programmer needs to express how packet headers are to be processed.  For example, TTL fields must be decremented and tested, new tunnel headers may need to be
added, and checksums may need to be computed. This motivates P4's use of an imperative control flow program to describe header field processing
using the declared header types and a primitive set of actions.

We could use a language such as Click~\cite{click}, which builds switches from
modules composed of arbitrary C++.  Click is extremely expressive, and very suitable for expressing how packets are processed in the kernel of a CPU. But it is insufficiently constrained for our needs---we need a language that mirrors the parse-match-action pipelines in dedicated hardware.  In addition, Click is not designed for a controller-switch architecture and hence does not allow programmers to describe \matchaction tables that are dynamically populated by well-typed rules.  Finally, Click makes it difficult to infer dependencies that constrain parallel execution---as we now discuss.

\begin{figure}[h]
\centering
\includegraphics[width=\columnwidth]{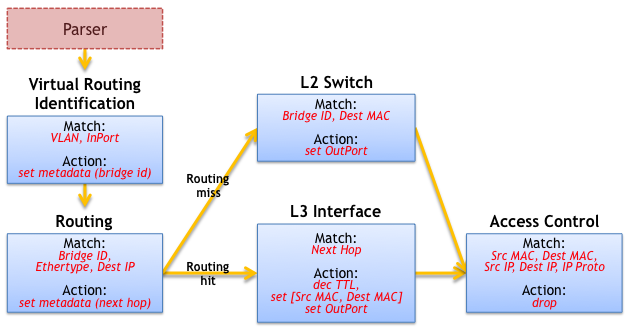}
\caption{Table dependency graph for an L2/L3 switch.}
\label{fig:tdg}
\end{figure}

A packet processing language must allow the programmer to express (implicitly or explicitly) any serial dependencies between header fields. Dependencies determine which tables can be executed in parallel. For example, sequential execution is required for an IP routing table and an ARP table due to the data dependency between them. Dependencies can be identified by analyzing {\em Table Dependency Graphs} (TDGs); these graphs describe the field inputs, actions, and control flow between tables. Figure~\ref{fig:tdg} shows an example table dependency graph for an L2/L3 switch. TDG nodes map directly to \matchaction tables, and a dependency analysis identifies where each table may reside in the pipeline. Unfortunately TDGs are not readily accessible to most programmers; programmers tend to think of packet processing algorithms using imperative constructs rather than graphs.

This leads us to propose a two-step compilation process.  At the highest level, programmers express packet processing programs using an imperative language representing the control flow (P4);  below this, a compiler translates the P4  representation to TDGs to facilitate dependency analysis, and then maps the TDG to a specific switch target.  P4 is designed to make it easy
to translate a P4 program into a TDG.   In summary, P4 can be considered to be a sweet  spot between the generality of say Click (that makes it difficult to infer dependencies and map to hardware) and the inflexibility of OpenFlow 1.0 (that makes it impossible to reconfigure protocol processing).

\begin{comment}
A {\em table graph} or {\em table dependency graph} follows naturally from the
abstract model description and is a potential configuration language. Such a
graph would express the desired tables, their logical arrangement, their
inputs, and the actions that each may perform. A compiler can identify
dependencies between tables by analyzing the graph, and then manipulate the
graph to map it onto a switch's resources.  (A description of the packet
formats is also necessary.)

For many programmers, table graphs are likely to be an awkward mechanism for
expressing packet processing algorithms; instead they are likely to be
accustomed to using imperative languages to express algorithms. As such, our P4
language contains imperative and declarative elements: the programmer expresses
packet formats and switch tables using declarative constructs, and expresses
processing algorithms using imperative constructs.

Although programmers may not work directly with table dependency graphs, the P4
compiler uses table dependency graphs as an intermediate representation. Table
dependency graphs are switch-agnostic, and can be manipulated and translated by
the compiler to the resources available within the given switch.
\end{comment}

\section{P4 Language By Example}

%We introduce P4 by means of an example.

We explore P4 by examining a simple example in-depth.
Many network deployments differentiate between an edge and a core;
end-hosts are directly connected to edge devices, which are in turn
interconnected by a high-bandwidth core. Entire protocols have been
designed to support this architecture (such as MPLS~\cite{MPLS} and
PortLand~\cite{PortLand}), aimed primarily at simplifying forwarding
in the core.

Consider an example L2 network deployment with top-of-rack (ToR)
switches at the edge connected by a two-tier core. We will assume the
number of end-hosts is growing and the core L2 tables are overflowing.
MPLS is an option to simplify the core, but implementing a label
distribution protocol with multiple tags is a daunting task.
PortLand looks interesting but requires rewriting MAC addresses---possibly
breaking existing network debugging tools---and requires new agents to respond
to ARP requests.

P4 lets us express a custom solution with minimal changes to the
network architecture.  We call our toy example \emph{mTag}: it
combines the hierarchical routing of PortLand with simple MPLS-like
tags. The routes through the core are encoded by a 32-bit tag composed
of four single-byte fields.  The 32-bit tag can carry a ``source
route'' or a destination locator (like PortLand's Pseudo MAC). Each
core switch need only examine one byte of the tag and switch on that
information.  In our example, the tag is added by the first ToR
switch, although it could also be added by the end-host NIC.

The {\em mTag} example is intentionally very simple to focus our attention on the P4 language. The P4 program for an entire switch would be many times more complex in practice.

\begin{comment}
\subsection{Underlying Hardware Assumptions}

P4 is intended to be target-independent so that one
P4 program can be compiled to switches supplied by multiple different
vendors.  Compliant hardware platforms will have to 
satisfy some basic requirements.  

First, the switch must support two modes of execution: (1) A
configuration mode in which information about packet formats and the
structure of \matchaction tables is communicated to the switch for
planning purposes, and (2) a population mode in which rules conforming
to the specifications are added and removed from the
tables.  Second, to implement P4 in its full generality,
it must be possible to configure the hardware's
packet parser to identify and extract new fields from a packet.
Third, tables within the target must support matching of all defined fields.
Fourth, the target must support implementation of a range of
protocol-independent packet-processing primitives, including copying,
addition, removal, and modification of both old and new fields as well
as metadata.

This model makes more requirements of the underlying hardware than 
conventional OpenFlow.  In particular, OpenFlow assumes a fixed parser, 
whereas our model supports a programmable parser that allows new headers to 
be defined. OpenFlow assumes the \matchaction tables are laid
out in sequence whereas we support both sequential and parallel processing
units.  Finally, we require actions to be defined using reusable,
protocol-independent primitives.
\end{comment}

\subsection{P4 Concepts}

A P4
program contains definitions of the following key components:

\begin{itemize}
\item{\bf Headers:} A header definition describes the sequence and
structure of a series of fields.  It includes specification of field widths
and constraints on field values.
\item{\bf Parsers:} A parser definition specifies how to identify headers and
  valid header sequences within packets.
\item{\bf Tables:} \Matchaction tables are the mechanism for performing
packet processing.  The P4 program defines the fields 
on which a table may match and the actions it may execute.
\item{\bf Actions:} P4 supports construction of complex actions from
simpler protocol-independent primitives.  These
complex actions are available within \matchaction tables.
\item{\bf Control Programs:} The control program determines the order
  of \matchaction tables that are applied to a packet.  A simple
  imperative program describe the flow of control between
  \matchaction tables.
\end{itemize}

\noindent
Next, we show how each of these components
contributes to the definition of an idealized \emph{mTag} processor in P4.

\subsection{Header Formats}
A design begins with the specification of header formats.  Several
domain-specific languages have been proposed for 
this~\cite{packettypes,datascript,fisher+:pads}; P4 borrows a number of
ideas from them.  In general, each header is specified by declaring 
an ordered list of field names together with their widths. Optional field
annotations allow constraints on value ranges or maximum lengths for
variable-sized fields.
For example, standard Ethernet and VLAN headers are specified as follows:
{\small
\begin{verbatim}
header ethernet {
    fields {
        dst_addr : 48; // width in bits
        src_addr : 48;
        ethertype : 16;
    }
}

header vlan {
    fields {
        pcp : 3;
        cfi : 1;
        vid : 12;
        ethertype : 16;
    }
}
\end{verbatim}
}

The {\em mTag} header can be added without altering existing declarations.
The field names indicate that the core has two
layers of aggregation.  Each core switch is programmed with rules to
examine one of these bytes determined by its location in the hierarchy
and the direction of travel (up or down).

{\small
\begin{verbatim}
header mTag {
    fields {
        up1 : 8;
        up2 : 8;
        down1 : 8;
        down2 : 8;
        ethertype : 16;
    }
}
\end{verbatim}
}

\subsection{The Packet Parser}
\label{sec:mtag_parser}
P4 assumes the underlying switch can implement 
a state machine that traverses packet headers from start 
to finish, extracting field values as it goes. The extracted field values are sent to the \matchaction tables for processing.

\begin{comment}
In general, a value in the packet header determines the next 
header to extract, e.g., the L2 Ethertype or L3 ProtocolID 
field.  P4 also supports other ways to determine the next header, as described 
in~\cite{Parse}.
\end{comment}

P4 describes this state machine directly as the set of transitions
from one header to the next.  Each transition may be triggered 
by values in the current header. For example, we describe the 
{\em mTag} state machine as follows. 

{\small
\begin{verbatim}
parser start{
    ethernet;
}

parser ethernet {
    switch(ethertype) {
        case 0x8100: vlan;
        case 0x9100: vlan;
        case  0x800: ipv4;
        // Other cases
    }
}

parser vlan {
    switch(ethertype) {
        case 0xaaaa: mTag;
        case  0x800: ipv4;
        // Other cases
    }
}

parser mTag {
    switch(ethertype) {
        case 0x800: ipv4;
        // Other cases
    }
}
\end{verbatim}
}

Parsing starts in the {\tt start} state and proceeds until an explicit
{\tt stop} state is reached or an unhandled case is encountered (which
may be marked as an error).  Upon reaching a state for a new header,
the state machine extracts the header using its specification and
proceeds to identify its next transition.  The extracted headers are
forwarded to \matchaction processing in the back-half of the switch
pipeline.
%The specification of a header name in the
%parser control flow indicates that the header is extracted, the
%packet location pointer is updated and control is transfered to the
%corresponding state.

The parser for {\em mTag} is very simple: it has only four
states. Parsers in real networks require many more states; for
example, the parser defined by Gibb \emph{et. al.}~\cite[Figure 3(e)]
{Parse} expands to over one hundred states.
%This illustrates the need for an automated tool chain to
%compile P4 programs to different targets. \dpw{this commented out sentence
% is not implied by the prior sentence.  Having a large number of parser
% states does not imply one needs to compile P4 to different targets ...?}

\subsection{Table Specification}

Next, the programmer describes how the defined header fields are to 
be matched in the \matchaction stages (e.g., should they be exact 
matches, ranges, or wildcards?) and what actions should be performed 
when a match occurs. 

In our simple {\em mTag} example, the edge switch matches on the 
L2 destination and VLAN ID, and selects an {\em mTag} to add to the 
header.  The programmer defines a table to match on 
these fields and apply an action to add the {\em mTag} 
header (see below). 
The {\tt reads} attribute declares which fields to match, qualified 
by the match type (exact, ternary, etc).  
The {\tt actions} attribute lists the possible actions which may
be applied to a packet by the table.  Actions are explained in the
following section.  The {\tt max\_size} attribute specifies how 
many entries the table should support. 

The table specification allows a compiler to decide how much 
memory it needs, and the memory type (e.g., TCAM or SRAM) to implement 
the table.

{\small
\begin{verbatim}
table mTag_table {
    reads {
        ethernet.dst_addr : exact;
        vlan.vid : exact;
    }
    actions {
        // At runtime, entries are programmed with params
        // for the mTag action.  See below.
        add_mTag;
    }
    max_size : 20000;
}
\end{verbatim}
}

For completeness and for later discussion, we present brief
definitions of other tables that are referenced by the Control Program (\S\ref{control-flow}).

{\small
\begin{verbatim}
table source_check {
    // Verify mtag only on ports to the core
    reads {
        mtag : valid; // Was mtag parsed?
        metadata.ingress_port : exact;
    }
    actions { // Each table entry specifies *one* action

        // If inappropriate mTag, send to CPU
        fault_to_cpu;

        // If mtag found, strip and record in metadata
        strip_mtag;

        // Otherwise, allow the packet to continue
        pass;
    }
    max_size : 64; // One rule per port
}

table local_switching {
    // Reads destination and checks if local
    // If miss occurs, goto mtag table.
}

table egress_check {
    // Verify egress is resolved
    // Do not retag packets received with tag
    // Reads egress and whether packet was mTagged
}
\end{verbatim}
}

\subsection{Action Specifications}

P4 defines a collection of primitive actions from 
which more complicated actions are built. Each P4 program declares a set of
action functions that are composed of action primitives; these action functions
simplify table specification and population. P4 assumes parallel execution of
primitives within an action function. (Switches incapable of parallel execution
may emulate the semantics.)

The \texttt{add\_mTag} action referred to above
is implemented as follows:

{\small 
\begin{verbatim}
action add_mTag(up1, up2, down1, down2, egr_spec) {
    add_header(mTag);
    // Copy VLAN ethertype to mTag
    copy_field(mTag.ethertype, vlan.ethertype);
    // Set VLAN's ethertype to signal mTag
    set_field(vlan.ethertype, 0xaaaa);
    set_field(mTag.up1, up1);
    set_field(mTag.up2, up2);
    set_field(mTag.down1, down1);
    set_field(mTag.down2, down2);

    // Set the destination egress port as well
    set_field(metadata.egress_spec, egr_spec);
}
\end{verbatim}
}

If an action needs parameters (e.g., the {\tt up1} value for the 
{\em mTag}), it is supplied from the match table at runtime.

In this example, the switch inserts the {\em mTag} after the VLAN tag,
copies the VLAN tag's ethertype into the {\em mTag} to indicate what
follows, and sets the VLAN tag's ethertype to \texttt{0xaaaa} to
signal {\em mTag}.  Not shown are the inverse action specification that strips an {\em mTag} from a packet and the table to apply this action in edge switches.

P4's primitive actions include:

\begin{itemize}
\item{\tt set\_field:}  Set a specific field in a header to a value. Masked sets are supported.
\item{\tt copy\_field:}  Copy one field to another.
\item{\tt add\_header:}  Set a specific header instance (and all its fields) as valid.
\item{\tt remove\_header:}  Delete (``pop'') a header (and all its fields) from a packet.
\item{\tt increment:} Increment or decrement the value in a field.
\item{\tt checksum:} Calculate a checksum over some set of header fields (e.g., an IPv4 checksum).
\end{itemize}

\noindent
We expect most switch implementations will restrict action processing
to permit only header modifications that are consistent with the
specified packet format.

\subsection{The Control Program}
\label{control-flow}
Once tables and actions are defined, the only remaining task is to specify
the flow of control from one table to the next.  Control flow is
specified as a program via a collection of functions, conditionals,
and table references.

\begin{figure}[h]
\centering
\includegraphics[width=3in]{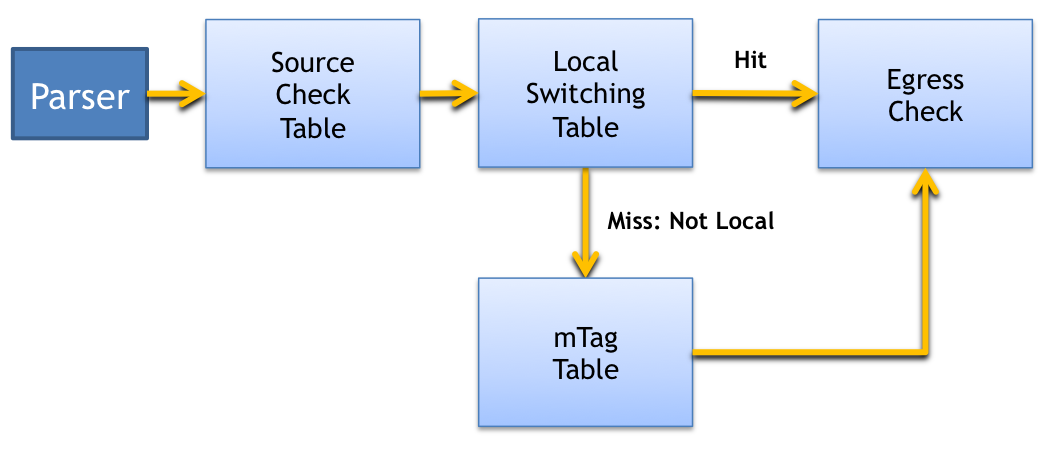}
\caption{Flow chart for the {\em mTag} example.}
\label{fig:mtag_flow}
\end{figure}

Figure~\ref{fig:mtag_flow} shows a graphical representation of
the desired control flow for the {\em mTag} 
implementation on edge switches.
After parsing, the {\tt source\_check} table verifies consistency between
the received packet and the ingress port.  For example, {\em mTags} should
only be seen on ports connected to core switches.  The {\tt source\_check} also
strips {\em mTags} from the packet, recording whether the packet had
an {\em mTag} in metadata.  Tables later in the pipeline may match on this
metadata to avoid retagging the packet.

The {\tt local\_switching} table is then executed. If this table ``misses,'' it
indicates that the packet is not destined for a locally connected
host.  In that case, the {\tt mTag\_table} (defined above) is applied to
the packet.  Both local and core forwarding control can be 
processed by the {\tt egress\_check}
table which handles the case of an unknown destination by sending
a notification up the SDN control stack.  

The imperative representation of this packet processing pipeline
is as follows:

{\small 
\begin{verbatim}
control main() {
    // Verify mTag state and port are consistent
    table(source_check);

    // If no error from source_check, continue
    if (!defined(metadata.ingress_error)) {
        // Attempt to switch to end hosts
        table(local_switching);
    
        if (!defined(metadata.egress_spec)) {
            // Not a known local host; try mtagging
            table(mTag_table);
        }
    
        // Check for unknown egress state or
        // bad retagging with mTag.
        table(egress_check);
    }
}
\end{verbatim}
}

\begin{comment}
% This is an additional action referenced in the tables
\begin{verbatim}
action strip_mtag() {
    // Strip the tag from the packet...
    remove_header(mtag);
    // but keep state that it was mtagged.
    set_field(metadata.was_mtagged, 1);
}
\end{verbatim}
\end{comment}

%\input{example}
\section{Compiling a P4 Program}

For a network to implement our P4 program, we need a compiler to map
the target-independent description onto the target switch's specific
hardware or software platform. Doing so involves allocating the
target's resources and generating appropriate configuration for the
device.

\subsection{Compiling Packet Parsers}
For devices with \emph{programmable} parsers, the compiler translates the parser description into a parsing state machine, while for fixed parsers, the compiler merely verifies that the parser description is consistent with the target's parser. Details of generating a state machine and state table entries can be found in~\cite{Parse}.

Table~\ref{tbl:parser_state} shows state table entries for the \texttt{vlan}
and \texttt{mTag} sections of the parser (\S\ref{sec:mtag_parser}). Each entry
specifies the current state, the field value to match, and the next state. Other columns are omitted for brevity.

\begin{table}[H]
\begin{center}
\small
\begin{tabular}{ccc} %\hline
{\bf Current State} &
{\bf Lookup Value} &
{\bf Next State} \\ \hline

\texttt{vlan} & \texttt{0xaaaa} & \texttt{mTag} \\
\texttt{vlan} & \texttt{0x800} & \texttt{ipv4}  \\
\texttt{vlan} & \texttt{*} & \texttt{stop} \\
\texttt{mTag} & \texttt{0x800} & \texttt{ipv4}  \\
\texttt{mTag} & \texttt{*} & \texttt{stop}  \\
\end{tabular}
\end{center}
\caption{Parser state table entries for the \emph{mTag} example.}
\label{tbl:parser_state}
\end{table}

\subsection{Compiling Control Programs}
\label{compiling-flow}

The imperative control-flow representation in \S\ref{control-flow} is
a convenient way to specify the logical forwarding behavior of a
switch, but does not explicitly call out dependencies between tables
or opportunities for concurrency. We therefore employ a compiler to
analyze the control program to identify dependencies and look for
opportunities to process header fields in parallel.  Finally, the
compiler generates the target configuration for the switch. There are
many potential targets: for example, a software switch~\cite{OVS}, a
multicore software switch~\cite{CuckooConextPaper}, an
NPU~\cite{EZchip}, a fixed function switch~\cite{BRCMtrident}, or a
reconfigurable match table (RMT)
pipeline~\cite{forwarding-metamorphosis}.

As discussed in \S\ref{prog-lang}, the compiler follows a two-stage
compilation process. It first converts the P4 control program into an
intermediate {\em table dependency graph} representation which it analyzes to
determine dependencies between tables. A target-specific back-end then maps
this graph onto the switch's specific resources.

We briefly examine
how the {\em mTag} example would be implemented in different kinds of
switches:

{\bf Software switches:} A software switch provides complete
flexibility: the table count, table configuration, and parsing are
under software control. The compiler directly maps the mTag table
graph to switch tables. The compiler uses table type information to
constrain table widths, heights, and matching criterion (e.g., exact,
prefix, or wildcard) of each table. The compiler might also optimize
ternary or prefix matching with software data structures.

{\bf Hardware switches with RAM and TCAM:} A compiler can configure
hashing to perform efficient exact-matching using RAM, for the
\texttt{mTag\_table} in edge switches.  In contrast, the core mTag
forwarding table that matches on a subset of tag bits would be mapped
to TCAM.

{\bf Switches supporting parallel tables:}
The compiler can detect data dependencies and arrange tables in parallel or
in series. In the mTag example, the tables \texttt{local\_switching} and
\texttt{mTag\_table} can execute in parallel up to the execution of the action
of setting an mTag.

{\bf Switches that apply actions at the end of the pipeline:}
For switches with action processing only at the end of a pipeline, the compiler
can tell intermediate stages to generate metadata that is used to perform the
final writes.  In the mTag example, whether the mTag is added or removed could
be represented in metadata.

{\bf Switches with a few tables:} The compiler can map a large number
of P4 tables to a smaller number of physical tables.  In the mTag
example, the local switching could be combined with the mTag table.
When the controller installs new rules at runtime, the compiler's rule
translator can ``compose'' the rules in the two P4 tables to generate
the rules for the single physical table.

\section{Conclusion}
\label{conclusion}
The promise of SDN is that a single control plane can directly control a whole
network of switches. OpenFlow supports this goal by providing a single,
vendor-agnostic API. However, today's OpenFlow targets {\em fixed}-function
switches that recognize a pre-determined set of header fields and that process packets using a small set of predefined actions. The control plane cannot express how packets {\em should} be processed to best meet the needs of control applications.

We propose a step towards more flexible switches whose functionality is
specified---and may be changed---in the field.  The programmer decides how the
forwarding plane processes packets without worrying about implementation
details. A compiler transforms an imperative program into a table dependency graph that can be mapped to many specific target switches, including optimized hardware implementations.

We emphasize that this is only a first step, designed as a straw-man proposal for OpenFlow 2.0 to contribute to the debate. In this proposal, several aspects of a switch remain undefined (e.g., congestion-control primitives, queuing disciplines, traffic monitoring). However, we believe the approach of having a configuration language---and compilers that generate low-level configurations for specific targets---will lead to future switches that provide greater flexibility, and unlock the potential of software defined networks.

%\paragraph*{Acknowledgments}
%This work is supported in part by ONR grant N00014-12-1-0757, NSF
%grants CNS 1111520 and SHF 1016937, and a Google Research Award. Any
%opinions, findings, and recommendations are those of the authors and
%do not necessarily reflect the views of the NSF, ONR, or Google.

\bibliographystyle{ieeetr}
\bibliography{paper}

\end{document}